\documentclass{article}%
\usepackage{amsmath}
\usepackage{amsfonts}
\usepackage{amssymb}
\usepackage{graphicx}%
\providecommand{\U}[1]{\protect\rule{.1in}{.1in}}
\newtheorem{theorem}{Theorem}

\newtheorem{corollary}[theorem]{Corollary}

\newtheorem{proposition}[theorem]{Proposition}
\newtheorem{remark}[theorem]{Remark}

\begin{document}

\begin{center}
\textbf{New robust confidence intervals for the mean under dependence}

\bigskip

Martial Longla and Magda Peligrad

\end{center}

Department of Mathematics, University of Mississippi, University, MS 38677,
USA. E-mail: mlongla@olemiss.edu

and

Department of Mathematical Sciences, University of Cincinnati, PO Box 210025,
Cincinnati, Oh 45221-0025, USA. Email:peligrm@ucmail.uc.edu


\bigskip

\begin{center}
\bigskip

Abstract
\end{center}

The goal of this paper is to indicate a new method for constructing normal
confidence intervals for the mean, when the data is coming from stochastic
structures with possibly long memory, especially when the dependence structure
is not known or even the existence of the density function. More precisely we
introduce a random smoothing suggested by the kernel estimators for the
regression function. Applications are presented to linear processes and
reversible Markov chains with long memory.

\begin{center}

\end{center}

\section{Introduction and results}

Let us suppose that we have a stationary and ergodic sequence $(Y_{i})_{i\in
Z}$ with finite variance ($\mathrm{var}(Y_{0})=\sigma_{Y}^{2}<\infty)$. Denote
by $\mu_{Y}=EY_{0},$ the expected value of $Y.$ Also, denote as usual the
sample mean by%
\[
\bar{Y}_{n}=\frac{1}{n}\sum_{i=1}^{n}Y_{i}=\frac{1}{n}S_{n}^{Y}.
\]
By the Birkhoff ergodic theorem it is well-known that
\[
\lim_{n\rightarrow\infty}\bar{Y}_{n}=\mu_{Y}.
\]
If the sequence $(Y_{i})_{i\in Z}$ has short range dependence and we have
additional information on its dependence structure, such as martingale-like
conditions or mixing conditions, we can derive a central limit theorem for
$\sqrt{n}(\bar{Y}_{n}-\mu_{Y}),$ which naturally leads to the construction of
confidence intervals for $\mu_{Y}$ based on normal distribution scalars.
Without other information on the dependence structure of $(Y_{i})_{i\in Z}$,
obviously, such a sequence might not obey the central limit theorem, and this
method is not possible to use. In this note we indicate a way to construct
normal confidence intervals for $\mu_{Y}$ based on a smoothing method inspired
by Nadaraya-Watson estimators.

For the purpose of this paper, we shall say that a sequence $(Y_{i})_{i\in Z}$
has long range dependence if $\mathrm{var}(S_{n}^{Y})/n\rightarrow\infty$ and
short range if $\mathrm{var}(S_{n}^{Y})$ behaves linearly in $n.$

Given a sample $(X_{i},Y_{i})_{1\leq i\leq n}$ from a random vector $(X,Y)$ on
a probability space $(\Omega,K,P)$, the well-known Nadaraya-Watson estimator
(see Nadaraya (1964)\ and Watson (1964), or pages 126-127 in H\"{a}rdle
(1991)) is defined by%
\[
\hat{m}_{n}(x)=\frac{1}{nh_{n}\hat{f}_{n}(x)}\sum_{i=1}^{n}Y_{i}K(\frac
{1}{h_{n}}(X_{i}-x)),
\]
where%
\[
\hat{f}_{n}(x)=\frac{1}{nh_{n}}\sum_{i=1}^{n}K(\frac{1}{h_{n}}(X_{i}-x)).
\]
This estimator has been widely studied in the literature. For instance, when
the vector $(X,Y)$ has joint density $f(x,y)$ say, $\hat{m}_{n}(x)$ is used to
estimate
\[
E(Y|X=x)=r(x)=\int y[f(x,y)/f(x)]dy.
\]
Furthermore, when $K$ is a kernel with several properties, $h_{n}$ is a
sequence of positive numbers (bandwidth) such that
\begin{equation}
h_{n}\rightarrow0\text{ and }nh_{n}\rightarrow\infty\text{ as }n\rightarrow
\infty, \label{band condition}%
\end{equation}
under various smoothness assumptions on $(X,Y)$ and various\ dependence
assumptions on the process $(X_{i},Y_{i})_{i\in Z}$, the speed of convergence
of $\hat{m}_{n}(x)\ $to $r(x)$ was pointed out in numerous papers. The
dependence structure considered in the literature is rather restrictive, of
the weak dependence type, such as mixing conditions, function of mixing
sequences or martingale-like conditions. We mention for instance results in
Bradley (1983), Collomb (1984), Peligrad (1992), Yoshihara (1994), Bosq
(1996), Bosq et.al. (1999), Long and Qian (2013), and Hong and Linton (2016)
among many others.

Now, let us notice that if the variables $(X_{i})_{i\in Z}$ are independent of
$(Y_{i})_{i\in Z},$ we have $E(Y|X)=E(Y)=\mu_{Y}.$ Inspired by Nadaraya-Watson
estimator, the goal of our paper is to indicate how this observation can be
used to develop robust procedures for constructing normal interval estimates
for the mean $\mu_{Y}$ by using the estimator%
\[
\hat{m}_{n}(0)=\frac{1}{nh_{n}\hat{f}_{n}(0)}\sum_{i=1}^{n}Y_{i}K(\frac
{1}{h_{n}}X_{i}),
\]
when there is very little information about the dependence structure of the
sequence $(Y_{i})_{i\in Z}$ or the existence of the density of $Y.$

The procedure we propose is the following. The data $(Y_{i})_{1\leq i\leq n}$
consists of a sample from a stationary and ergodic sequence $(Y_{i})_{i\in
Z}.$ Independently of $(Y_{i})_{1\leq i\leq n}$ we generate a random sample
$(X_{i})_{1\leq i\leq n}$, from a distribution with bounded density $f(x),$
continuous at the origin, with $f(0)\neq0$. It is known that $\hat{f}_{n}(0)$
is an asymptotically unbiased estimator for $f(0),\ $provided the bandwidths
$h_{n}$ satisfies the condition (\ref{band condition})\ and the kernel $K$
satisfies (see Parzen, 1962 or H\"{a}rdle, page 59) the following condition
\begin{equation}
K\text{ is a symmetric bounded density function.}\label{condition K}%
\end{equation}
Under these conditions, $\lim_{n\rightarrow\infty}\hat{f}_{n}(0)=f(0)$ in
$L_{2}$. Therefore, by Slutsky's theorem, we can replace the study of the
limiting distribution of $\hat{m}_{n}$ by that of its asymptotic equivalent
estimator%
\begin{equation}
\hat{r}_{n}=\frac{1}{nh_{n}f(0)}\sum_{i=1}^{n}Y_{i}K(\frac{1}{h_{n}}%
X_{i}).\label{def prime}%
\end{equation}
The estimator $\hat{m}_{n}(0)$ is an unbiased estimator of $\mu_{Y},$ while
$\hat{r}_{n}$ is asymptotically unbiased. However $\hat{m}_{n}(0)$ has the
disadvantage that it introduces an error due to replacing the known quantity
$f(0)\ $by its estimate $\hat{f}_{n}(0)$. In addition $\hat{r}_{n}$ is easier
to analyze. This is the reason why we prefer to use $\hat{r}_{n}$ as our
proposed estimator for $\mu_{Y}.$ We shall provide a central limit theorem, a
functional central limit theorem and also discuss the optimal bandwidth which
minimizes the mean square error.

To establish these results, we use the independence structure of the smoothing
sequence $(X_{i})_{i\in Z}$ that allows us not to restrict the dependence
structure of $(Y_{i})_{i\in Z}$ and also not to impose the existence of the
density of $Y$. The closest idea to this one is the block-wise bootstrap. For
instance in the paper by Peligrad (1998),\ the central limit theorem for the
mean is obtained via bootstrap smoothing, for a sequence that does not satisfy
the CLT, but rather satisfies some restrictive mixing conditions. In the
sequel we denote by $\Rightarrow$ the convergence in distribution. For
positive sequences of numbers $a_{n}=O(b_{n})$ means $\lim\sup_{n\rightarrow
\infty}a_{n}/b_{n}<\infty$; $a_{n}=o(b_{n})$ means $\lim_{n\rightarrow\infty
}a_{n}/b_{n}=0.$ We use the notation $a_{n}\sim b_{n}$ for $\lim
_{n\rightarrow\infty}a_{n}/b_{n}=1.$

Besides condition (\ref{band condition}) we shall impose the following
assumption on the bandwidths sequence $(h_{n})_{n\geq1}:$
\begin{equation}
\sqrt{nh_{n}}(\bar{Y}_{n}-\mu_{Y})\rightarrow^{P}0,\label{ratePP}%
\end{equation}
which is implied by
\begin{equation}
nh_{n}\mathrm{var}(\bar{Y}_{n})\rightarrow0.\label{rateP}%
\end{equation}
Note that we can always find a sequence $(h_{n})_{n\geq1}$ satisfying both
conditions (\textbf{\ref{band condition}})\ and (\ref{rateP}), provided that
$\mathrm{var}(\bar{Y}_{n})\rightarrow0$.

We shall establish the following theorem:

\begin{theorem}
\label{Theo1} Assume that $(Y_{i})_{i\in Z}$ is a stationary and ergodic
sequence with finite second moments and conditions
(\textbf{\ref{band condition}}) and condition (\ref{ratePP})\ are satisfied.
Also assume that $K$ satisfies condition (\ref{condition K}) and that
$(X_{i})_{i\in N}$ is an i.i.d. sequence of random variables, independent of
$Y,$ having a bounded density function $f(x),$ continuous at the origin, with
$f(0)\neq0$. Then we have
\[
\frac{\sqrt{nh_{n}}}{\sqrt{\overline{Y_{n}^{2}}}}(\hat{r}_{n}-\mu
_{Y})\Rightarrow N(0,\frac{1}{f(0)}\int K^{2}(x)dx).
\]
where $\ \overline{Y_{n}^{2}}=\sum_{i=1}^{n}Y_{i}^{2}/n$ and $\hat{r}_{n}$ is
defined by (\ref{def prime})$.$\textbf{ }
\end{theorem}

By combining this theorem with the consistency of $\hat{f}_{n}(0)$ we obtain that

\begin{corollary}
Under the conditions of Theorem \ref{Theo1} we also have%
\[
\frac{\sqrt{nh_{n}\hat{f}_{n}(0)}}{\sqrt{\overline{Y_{n}^{2}}}}(\hat{m}%
_{n}-\mu_{Y})\Rightarrow N(0,\int K^{2}(x)dx).
\]

\end{corollary}

Based on Theorem \ref{Theo1} we can construct confidence intervals for the mean:

\begin{corollary}
Under the conditions of Theorem \ref{Theo1}, for $0<\alpha<1,$ a
$(1-\alpha)100\%$ confidence interval for $\mu_{Y}$ is%
\begin{equation}
\left\{  \hat{r}_{n}-z_{\alpha/2}\left(  \frac{\overline{Y_{n}^{2}}\int
K^{2}(x)dx}{nh_{n}f(0)}\right)  ^{1/2},\text{ }\hat{r}_{n}+z_{\alpha/2}\left(
\frac{\overline{Y_{n}^{2}}\int K^{2}(x)dx}{nh_{n}f(0)}\right)  ^{1/2}\right\}
,\label{conf int}%
\end{equation}
where $P(-z_{\alpha/2}<Z<z_{\alpha/2})=1-\alpha$ and $Z$ is a standard normal variable.
\end{corollary}

Let us notice that, at no extra cost, our result can be also formulated as a
functional CLT. If we consider the stochastic process%
\[
\hat{r}_{n}(t)=\frac{1}{nh_{n}f(0)}\sum_{i=1}^{[nt]}Y_{i}K(\frac{1}{h_{n}%
}X_{i}),\text{ }\hat{m}_{n}(t)=\frac{1}{nh_{n}\hat{f}(0)}\sum_{i=1}%
^{[nt]}Y_{i}K(\frac{1}{h_{n}}X_{i}),
\]
from the proof of Theorem \ref{Theo1} and Donsker's theorem (see Theorem 8.2
in Billingsley, 1999) we obtain:

\begin{corollary}
Under the conditions of Theorem \ref{Theo1} we have%
\[
\sqrt{nh_{n}}(\hat{r}_{n}(t)-\mu_{Y})/\sqrt{\overline{Y_{n}^{2}}}%
\Rightarrow\left(  \frac{1}{f(0)}\int K^{2}(x)dx\right)  ^{1/2}W(t),
\]
where $W(t)$ is the standard Brownian motion, and also%
\[
\sqrt{nh_{n}\hat{f}_{n}(0)}(\hat{m}_{n}(t)-\mu_{Y})/\sqrt{\overline{Y_{n}^{2}%
}}\Rightarrow\left(  \int K^{2}(x)dx\right)  ^{1/2}W(t).
\]

\end{corollary}

Our paper is organized as follows: In Section 2 we prove Theorem \ref{Theo1}.
In Section 3 we discuss the data driven selection of the optimal bandwidth to
be used in confidence intervals. Several applications to processes with long
memory are given in Section 4. In the last section we mention several remarks.

\section{Proof of Theorem \ref{Theo1}}

For convenience, we shall drop the index $n$ from the notation of $h_{n}$.~We
condition on $(Y_{i})_{i\in Z}$ and we shall first find the limiting
distribution of a related sequence of random variables under the regular
conditional probability $P_{Y}^{\omega}(\cdot)=P(\cdot|(Y_{i})_{i\in
Z})(\omega)$. In the sequel $E_{Y}^{\omega}$ denotes the expected value with
respect to $P_{Y}^{\omega}.$ We introduce the sequence of random variables
\begin{equation}
Z_{n,i}=\frac{1}{\sqrt{h}}\left(  K(\frac{1}{h}X_{i})-E(K(\frac{1}{h}%
X_{i}))\right)  Y_{i}=X_{n,i}Y_{i}, \label{defZ}%
\end{equation}
where%
\[
X_{n,i}=\frac{1}{\sqrt{h}}\left[  K(\frac{1}{h}X_{i})-E(K(\frac{1}{h}%
X_{i}))\right]  .
\]
Note that, by the independence of sequences $(Y_{i})_{i\in Z}$ and
$(X_{i})_{i\in Z},$ for almost all $\omega,$ we have%
\[
E_{Y}^{\omega}(Z_{n,i})=Y_{i}(\omega)E(X_{n,i})=0.
\]
Denote%
\[
W_{n}=\frac{1}{\sqrt{n}}\sum_{i=1}^{n}Z_{n,i}=\frac{1}{\sqrt{n}}\sum_{i=1}%
^{n}X_{n,i}Y_{i}.
\]
Let us find the limiting distribution of $W_{n}$ under $P_{Y}^{\omega},$ for
almost all $\omega$. We start by constructing $\Omega^{\prime}$ such that, for
all $\omega\in\Omega^{\prime}$ the following convergences hold:%

\begin{equation}
\lim_{n\rightarrow\infty}\frac{1}{n}\sum_{i=1}^{n}Y_{i}^{2}(\omega)=E(Y^{2})
\label{limY}%
\end{equation}
and for all $A,$ positive integer%
\begin{equation}
\lim_{n\rightarrow\infty}\frac{1}{n}\sum_{i=1}^{n}Y_{i}^{2}(\omega
)I(|Y_{i}|(\omega)>A)=E[Y^{2}I(|Y|>A)]. \label{limYA}%
\end{equation}
This is possible because $(Y_{i})_{i\in Z}$ is ergodic, so the convergences in
(\ref{limY})\ and (\ref{limYA})\ hold on sets of measure $1$. We construct
$\Omega^{\prime}$ as a countable intersection of these sets, which will also
have measure $1$. Fix $\omega\in\Omega^{\prime}.$

Under $P_{Y}^{\omega}$, $(W_{n})_{n\geq1}$ becomes a sum of a triangular array
of independent random variables. Therefore, in order the establish the CLT, we
have to take care of the limiting variance and then verify the Lindeberg's
condition. All the integrals below are taken over $R=(-\infty,\infty).$

First we recall that for all $i\in N,$%

\begin{align*}
\mathrm{var}(X_{n,i})  &  =\frac{1}{h}\int K^{2}(\frac{t}{h})f(t)dt-\frac
{1}{h}(\int K(\frac{t}{h})f(t)dt)^{2}\\
&  =\int K^{2}(t)f(th)dt-h(\int K(t)f(th)dt)^{2}.
\end{align*}
So, by Bochner's theorem and condition (\ref{condition K}),
\begin{equation}
\lim_{n\rightarrow\infty}\ \mathrm{var}(X_{n,i})=\lim_{n\rightarrow\infty
}E(X_{n,i}^{2})=f(0)\int K^{2}(u)du=C_{1}. \label{varX}%
\end{equation}
By the independence of sequences $(Y_{i})_{i\in Z}$ and $(X_{i})_{i\in Z}$ and
stationarity we have%

\[
\sigma_{n}^{2}(\omega)=\mathrm{var}_{Y}^{\omega}(W_{n})=\frac{1}{n}\sum
_{i=1}^{n}Y_{i}^{2}(\omega)\mathrm{var}(X_{n,1})=\overline{Y_{n}^{2}}%
(\omega)\mathrm{var}(X_{n,1})
\]
and therefore, by (\ref{limY})%
\begin{equation}
\lim_{n\rightarrow\infty}\sigma_{n}^{2}(\omega)=\lim_{n\rightarrow\infty}%
\frac{C_{1}}{n}\sum_{i=1}^{n}Y_{i}^{2}(\omega)=C_{1}E(Y^{2})\text{.}%
\label{limvar}%
\end{equation}
Let us establish now the Lindeberg's condition under $P_{Y}^{\omega}$. 

Denote $\sigma_{n}(\omega)=\sqrt{\sigma_{n}^{2}(\omega)}.$ We have to show
that, for every $\varepsilon>0,$
\begin{equation}
\lim_{n\rightarrow\infty}\frac{1}{\sigma_{n}(\omega)}\frac{1}{n}\sum_{i=1}%
^{n}E_{Y}^{\omega}[X_{n,i}^{2}Y_{i}^{2}I(|X_{n,i}Y_{i}|\geq\varepsilon
\sigma_{n}(\omega)\sqrt{n})]=0.\label{Lind}%
\end{equation}
Now, by (\ref{limvar}) there is $N(\varepsilon,\omega)$ such that for all
$n>N(\varepsilon,\omega)$ we have $\sigma_{n}(\omega)\geq C_{1}E(Y^{2})/2.$ By
this remark, by the independence of the two sequences (see Example 33.7 in
Billingsley) and stationarity, we obtain%
\[
E_{Y}^{\omega}[X_{n,i}^{2}Y_{i}^{2}I(|X_{n,i}Y_{i}|\geq\varepsilon\sigma
_{n}(\omega)\sqrt{n})]=Y_{i}^{2}(\omega)E[X_{n,1}^{2}I(|X_{n,1}Y_{i}%
(\omega)|\geq\varepsilon\sigma_{n}(\omega)\sqrt{n})].
\]
It follows that, in order to show (\ref{Lind}),\ we have to show instead
\[
\lim_{n\rightarrow\infty}\frac{1}{n}\sum_{i=1}^{n}Y_{i}^{2}(\omega
)E[X_{n,1}^{2}I(|X_{n,1}Y_{i}(\omega)|\geq\varepsilon^{\prime}\sqrt{n})]=0,
\]
where we denoted $\varepsilon^{\prime}=\varepsilon C_{1}E(Y^{2})/2.$ Denote
the expression above:%
\[
G_{n}(\omega)=\frac{1}{n}E[X_{n,1}^{2}\sum_{i=1}^{n}Y_{i}^{2}(\omega
)I(|X_{n,1}Y_{i}(\omega)|\geq\varepsilon^{\prime}\sqrt{n})].
\]
We shall decompose the sum in two parts. Let $A$ be a positive integer and
define the index sets
\[
I_{1}(\omega)=(i:1\leq i\leq n,\text{ }|Y_{i}|(\omega)\leq A),
\]%
\[
I_{2}(\omega)=(i:1\leq i\leq n,\text{ }|Y_{i}|(\omega)>A).
\]
Note $\{1,2,...,n\}=I_{1}(\omega)\cup I_{2}(\omega).$ We write $\sum_{i=1}%
^{n}=\sum_{i\in I_{1}(\omega)}+\sum_{i\in I_{1}(\omega)}$ and, by using the
stationarity assumption, we shall upper bound $F_{n}$ in the following way:
\begin{equation}
G_{n}(\omega)\leq A^{2}E[X_{n,1}^{2}I(|X_{n,1}|\geq A^{-1}\varepsilon^{\prime
}\sqrt{n})]+E(X_{n,1}^{2})\frac{1}{n}\sum_{i=1}^{n}Y_{i}^{2}(\omega
)I(|Y_{i}|(\omega)>A).\label{Fn}%
\end{equation}
Note that :%
\begin{gather*}
E[X_{n,1}^{2}I(|X_{n,1}|\geq A^{-1}\varepsilon^{\prime}\sqrt{n})]=\frac{1}%
{h}\int K^{2}(\frac{v}{h})I(K(\frac{v}{h})\geq A^{-1}\varepsilon^{\prime}%
\sqrt{nh})f(v)dv=\\
\ \int K^{2}(u)I(K(u)\geq A^{-1}\varepsilon^{\prime}\sqrt{nh})f(uh)du.
\end{gather*}
Since $nh\rightarrow\infty$ and $K$ is bounded, this limit is $0$ as
$n\rightarrow\infty.$ By passing to the limit in (\ref{Fn})\ with
$n\rightarrow\infty$ and by using (\ref{varX}), we easily obtain%
\[
\lim\sup_{n\rightarrow\infty}G_{n}(\omega)=C_{1}E[Y^{2}I(|Y|>A)].
\]
By letting $A\rightarrow\infty,$ and using the fact that $Y$ has finite second
moment, we get%
\[
\lim_{n\rightarrow\infty}G_{n}(\omega)=0\text{.}%
\]
Therefore, the Lindeberg's condition is satisfied under $P_{Y}^{\omega}$. By
all this considerations, we obtain that the following quenched central limit
theorem holds: for any fixed $\omega\in\Omega^{\prime}$%
\[
W_{n}\Rightarrow N(0,C_{1}E(Y^{2}))\text{ under }P_{Y}^{\omega}.\text{ }%
\]
This quenched CLT\ is a stronger form of CLT. After representing it in terms
of characteristic function we can integrate with respect to the measure $P$
and we obtain the annealed CLT, namely%
\begin{equation}
W_{n}\Rightarrow N(0,C_{1}E(Y^{2}))\text{ under }P.\label{CLT}%
\end{equation}
Now recall the definition of $Z_{n,i}=h^{-1/2}\left(  K(\frac{1}{h}%
X_{i})-E(K(\frac{1}{h}X_{i}))\right)  Y_{i}$. Let us also note that by
definition (\ref{def prime}),
\[
\frac{1}{\sqrt{nh}}\sum_{i=1}^{n}Y_{i}K(\frac{1}{h}X_{i})=\sqrt{nh}f(0)\hat
{r}_{n}.
\]
So we can rewrite%
\begin{equation}
W_{n}=\sqrt{nh}{\LARGE (}f(0)\hat{r}_{n}-\frac{1}{nh}\sum_{i=1}^{n}%
Y_{i}E[K(\frac{1}{h}X_{i})]{\LARGE ).}\label{1}%
\end{equation}
Note that, by the properties of $K,$
\[
\lim_{n\rightarrow\infty}\sqrt{nh}(\frac{1}{nh}\sum_{i=1}^{n}Y_{i}E(K(\frac
{1}{h}X_{i}))-f(0)\mu_{Y})=\lim_{n\rightarrow\infty}\sqrt{nh}f(0)(\frac{1}%
{n}\sum_{i=1}^{n}Y_{i}-\mu_{Y})\ .
\]
If we impose (\ref{rateP}),\ then%
\begin{equation}
\sqrt{nh}(\frac{1}{nh}\sum_{i=1}^{n}Y_{i}E(K(\frac{1}{h}X_{i}))-f(0)\mu
_{Y})\rightarrow^{P}0\label{2}%
\end{equation}
and, by Theorem 25.2\ in Billingsley (1995), we obtain%
\[
\sqrt{nh}f(0){\LARGE (}\hat{r}_{n}-\mu_{Y}{\LARGE )}\Rightarrow N(0,C_{1}%
E(Y^{2})).
\]
By the ergodic theorem and Slutski's theorem we obtain the desired result.
$\square$

\section{Data driven bandwidth selection.\textbf{ \label{section band}}}

The method we propose introduces new parameters, the bandwidth sequence
$(h_{n})_{n\geq1}$. There is a vast literature on the selection of $h_{n}$ for
kernel estimation of the density and for the Nadaraya-Watson estimator of a
regression, under independence or weak dependence assumptions. They can be
found in books, such as in Section 5.1.2 in H\"{a}rdle (1990) or in surveys,
such as Jones et. al. (1996). Our case deals with possible long dependence for
$(Y_{i})_{i\in Z}$ but it benefits from the independence of $(Y_{i})_{i\in Z}$
and $(X_{i})_{i\in Z}$ and also from the fact that we know $f(x)$. If we
impose additional conditions on the smoothness of $f(x)$ and $K(x),$ namely
$f(x)$ has a continuous and bounded second derivative and $K$ satisfies
condition (\ref{condition K}) and $\int u^{2}K(u)du<\infty,$ we can analyze
the optimal bandwidth by optimizing the main part of the mean square error
under the constraint (\ref{rateP}). We shall see that this selection depends
on the strength of dependence of $(Y_{i})_{i\in Z}.$ As a matter of fact we
shall prove that 

\begin{proposition}
Under the  conditions above, the optimal data driven bandwidth to be used in
the confidence intervals is
\begin{equation}
h_{o}=[\frac{f(0)B\overline{Y_{n}^{2}}}{n(f"(0)A)^{2}(\bar{Y}_{n})^{2}}%
]^{1/5}\text{ provided that }\mathrm{var}(\bar{Y}_{n})=o(n^{-4/5})\text{ and
}\mu_{Y}\neq0.\label{ho}%
\end{equation}

\end{proposition}

\textbf{Proof}. Denote $V_{n,i}=h^{-1}K(X_{i}/h).$ We shall compute first the
bias%
\[
\mathrm{Bias}(\hat{r}_{n})=E(\hat{r}_{n}-\mu_{Y})=\frac{\mu_{Y}}{f(0)}%
E(\hat{f}_{n}(0))-\mu_{Y}=\frac{\mu_{Y}}{f(0)}\mathrm{Bias}\hat{f}_{n}(0).
\]
The variance of the estimator is%

\begin{align*}
\mathrm{var}(\hat{r}_{n}) &  =E[\hat{r}_{n}-\frac{\mu_{Y}}{f(0)}E(\hat{f}%
_{n}(0))]^{2}\\
&  =E[\hat{r}_{n}-\frac{\mu_{Y}}{f(0)}\hat{f}_{n}(0)+\frac{\mu_{Y}}{f(0)}%
(\hat{f}_{n}(0)-E(\hat{f}_{n}(0))]^{2}\\
&  =E[\hat{r}_{n}-\frac{\mu_{Y}}{f(0)}\hat{f}_{n}(0)]^{2}+\frac{\mu_{Y}^{2}%
}{f^{2}(0)}\mathrm{var}\hat{f}_{n}(0)=I+II.
\end{align*}
A simple computation shows that the first term%
\begin{gather*}
I=\frac{1}{f^{2}(0)}E(\frac{1}{n}\sum\nolimits_{i=1}^{n}(Y_{i}-\mu_{Y}%
)V_{n,i})^{2}\\
=\frac{1}{f^{2}(0)n^{2}}[n\sigma_{Y}^{2}E(V_{n,i}^{2})+2\sum\limits_{1\leq
i<j\leq n}cov(Y_{i},Y_{j})(EV_{n,i})^{2}]\\
\frac{1}{f^{2}(0)n^{2}}[n\sigma_{Y}^{2}E(V_{n,i}^{2})+(EV_{n,1})^{2}%
(\mathrm{var}S_{Y}-n\sigma_{Y}^{2})]=\\
\frac{1}{f^{2}(0)n^{2}}[n\sigma_{Y}^{2}\mathrm{var}V_{n,i}+(EV_{n,1}%
)^{2}\mathrm{var}S_{Y}]\\
=\frac{1}{f^{2}(0)}[\sigma_{Y}^{2}\mathrm{var}\hat{f}_{n}(0)+(EV_{n,1}%
)^{2}\mathrm{var}\bar{Y}_{n}].
\end{gather*}
Therefore, by combining these estimates, the mean square error is%

\begin{align*}
\mathrm{MSE}(\hat{r}_{n}) &  =E(\hat{r}_{n}-\mu_{Y})^{2}=\mathrm{var}\hat
{r}_{n}+[\mathrm{Bias}(\hat{r}_{n})]^{2}\\
&  =\frac{1}{f^{2}(0)}[E(Y^{2})\mathrm{var}\hat{f}_{n}(0)+(EV_{n,1}%
)^{2}\mathrm{var}\bar{Y}_{n}+\mu_{Y}^{2}[\mathrm{Bias}\hat{f}_{n}(0)]^{2}.
\end{align*}
Under the assumption (\ref{rateP}),
\[
(EV_{n,1})^{2}\mathrm{var}\bar{Y}_{n}=o(\frac{1}{nh_{n}}).
\]
When $f(x)$ has a continuous and bounded second derivative, according to
formula (2.3.2) in H\"{a}rdle (1990) we have%
\[
\mathrm{Bias}\hat{f}_{n}(0)=\frac{h^{2}}{2}f"(0)A+o(h^{2})\text{ as
}n\rightarrow\infty,
\]
where%
\[
A=\int x^{2}K(x)dx.
\]
Also by formula (2.3.3) in the same book%
\[
\mathrm{var}\hat{f}_{n}(0)=\frac{1}{nh}Bf(0)+o(\frac{1}{nh})\text{ as
}n\rightarrow\infty,
\]
where
\[
B=\int K^{2}(x)dx.
\]
It follows that%
\[
\mathrm{MSE}(\hat{r}_{n})=\frac{1}{f^{2}(0)}[\frac{E(Y^{2})}{nh}%
Bf(0)+\frac{h^{4}}{4}\mu_{Y}^{2}(f"(0)A)^{2}+o(\frac{1}{nh})+o(h^{4})].
\]
In order to minimize it, we set $0$ the derivative with respect to $h$ of the
main part and obtain
\[
h_{o^{\prime}}=[\frac{f(0)BE(Y^{2})}{n(f"(0)A)^{2}\mu_{Y}^{2}}]^{1/5},
\]
provided $\mu_{Y}\neq0.$ Since the optimal $h_{o^{\prime}}$ depends on the
unknown parameters $E(Y^{2})$ and $\mu_{Y}^{2}\neq0,$ we shall replace them by
plug in estimators which are consistent because of the ergodicity of
$(Y_{n})_{n}$ we obtain (\ref{ho}).  $\square$

\begin{remark}
This $h_{o}$ was obtained by imposing condition (\ref{rateP}). At the same
time $h_{o}$ has to satisfy (\ref{rateP}), leading to the restriction
$\mathrm{var}(\bar{Y}_{n})=o(n^{-4/5})$. Otherwise, if $\lim\sup
_{n\rightarrow\infty}(n^{4/5}\mathrm{var}(\bar{Y}_{n}))\neq0$ the MSE\ is
minimized when $h_{n}$ is the largest possible satisfying
(\ref{band condition}) and (\ref{rateP}).
\end{remark}

\section{Applications to stationary sequences with long memory}

\textbf{Example 1}.\textbf{ Restriction on the covariance structure}. Let us
first point an example of a sequence where no restriction of the dependence
structure will be assumed, or the distribution of $Y,$ except ergodicity and a
mild restriction on the covariances.

For a stationary and ergodic sequence of random variables $(Y_{k})_{k\in Z}$
with finite second moment, let us assume that $|\mathrm{cov}(Y_{0},Y_{k})|\sim
C(k^{-\alpha})$ as $k\rightarrow\infty$ for $\alpha>0.$ For $0<\alpha\leq1,$
the covariances are not summable and $(Y_{k})_{k\in Z}$ has long memory. Note
first that we have
\begin{equation}
nh\mathrm{var}(\bar{Y}_{n})\leq2h_{n}\sum_{k=0}^{n}|\mathrm{cov}(Y_{0}%
Y_{k})|=O(h_{n}n^{-\alpha+1})\text{ as }n\rightarrow\infty. \label{ap band}%
\end{equation}
Therefore the condition (\ref{rateP}) of Theorem \ref{Theo1} holds as soon as
$h_{n}=o(n^{-1+\alpha})$ and our CLT applies$.$

We shall see that when $\alpha>0.8,$ Theorem \ref{Theo1} can be applied with
an optimal $h_{0}\sim C(n^{-1/5}),$ where $C$ is as in formula (\ref{ho}).
Indeed, for this range both (\ref{ap band}) and (\ref{ho})$\ $are satisfied.
If $0<\alpha\leq0.8,$ the $\mathrm{MSE}$ will converge to $0$ at a rate slower
than $n^{-\alpha}.$ On the other hand, if we have $|\mathrm{cov}(Y_{0}%
,Y_{k})|\sim C(\log k)^{-1}$ then $\sum_{k=0}^{n}|\mathrm{cov}(Y_{0}%
Y_{k})|=O(n/\log n).$ If we take $h_{n}=o(\log n/n)$ Theorem \ref{Theo1} still
can be applied. In this case the rate of convergence to $0$ of the
$\mathrm{MSE}$ is slower than $(\log n)^{-1}$ as $n\rightarrow\infty.$ This
shows that when the memory is very long the rates of convergence can be rather
slow, therefore a very large sample size might be necessary.

\bigskip

\textbf{Example 2. Long memory linear processes.} Let $(\xi_{j})_{j\in Z}$ be
an i.i.d. sequence of random variables, centered with finite second moments.
Let $(a_{j})_{j\in Z}$ be a sequence of constants We consider the linear
process
\begin{equation}
Y_{k}=\sum_{j=-\infty}^{\infty}a_{k-j}\xi_{j}. \label{ln}%
\end{equation}
Denote $S_{n}=\sum_{k=1}^{n}Y_{k}$. If $\sum_{i\in{\mathbb{Z}}}a_{i}%
^{2}<\infty$, (\ref{ln}) is well defined a.s. and in $L_{2}.$ We can write
$S_{n}=\sum_{i=-\infty}^{\infty}b_{ni}\xi_{i}$ with
\[
b_{ni}=a_{1-i}+\cdots+a_{n-i}%
\]
Using this notation we have $\mathrm{var}(S_{n})=\mathrm{var}(\xi_{0}^{2}%
)\sum_{i}b_{ni}^{2}$. Then $\mathrm{var}(\sqrt{nh}(S_{n}/n))=h_{n}n^{-1}%
\sum_{i}b_{ni}^{2}.$

If we assume (\ref{band condition}) and that $h_{n}n^{-1}\sum_{i}b_{ni}%
^{2}\rightarrow0$, then the conclusion of Theorem \ref{Theo1} holds.

As a particular example we consider the important case of causal long-memory
processes with
\[
a_{i}=[l(i+1)](1+i)^{-\alpha},\text{ }i\geq0,\text{ with }1/2<\alpha<1,\text{
and }a_{i}=0\text{ otherwise}.
\]
Here $l(\cdot)$ is a slowly varying function at infinite. These processes have
long memory because $\sum_{j\geq0}|a_{j}|=\infty.$

For this case, $var(\bar{Y}_{n})\sim\kappa_{\alpha}n^{1-2\alpha}\ell^{2}(n)$
(see for instance Relations (12) in Wang \textit{et al.} (2003)), where
$\kappa_{\alpha}$ is a positive constant depending on $\alpha$. Theorem
\ref{Theo1} can be applied as soon as $h_{n}=o(n^{2(1-\alpha)}\ell
^{2}(n))^{-1}$ as $n\rightarrow\infty.$ In the range $0.9<\alpha<1,$ $h_{o}$
can be taken as in (\ref{ho}). Otherwise the $\mathrm{MSE}$ will converge to
$0$ at a rate slower than $\ell^{2}(n)/n^{2\alpha-1}$ as $n\rightarrow\infty.$

This example covers the ARFIMA $(0,d,0)$\ processes (cf. Granger and Joyeux
(1980); Hosking (1981)), which play an important role in financial time series
modeling and application. As a special case, let $0<d<1/2$ and $B$ be the
backward shift operator with $B\varepsilon_{k}=\varepsilon_{k-1}$,
\[
X_{k}=(1-B)^{-d}\xi_{k}=\sum_{i\geq0}a_{i}\xi_{k-i},\text{ where }a_{i}%
=\frac{\Gamma(i+d)}{\Gamma(d)\Gamma(i+1)}.
\]
Here $\lim_{n\rightarrow\infty}a_{n}/n^{d-1}=1/\Gamma(d)$. For this case
$var(\bar{Y}_{n})\sim\kappa_{d}n^{2d-1}$ and condition (\ref{ho}) becomes
$h_{n}=o(n^{-2(1-d)}).$ For this case $h_{o}$ can taken as in (\ref{ho}) for
$0<d<0.1.$ For $0.1\leq d<1/2,$ and a selection of $h_{n}=o(n^{-2/d})$, the
$\mathrm{MSE}$ will converge to $0$ at a rate slower than the order $n^{2d-1}$
as $n\rightarrow\infty.$

\bigskip

\textbf{Example 3. A long memory reversible Markov chain. }For a nonlinear
example we would like to mention an example given in Zhao et al. (2010),
describing a stationary and ergodic reversible Markov chain, which does not
satisfy the CLT. This is their Example 2. Let $1<\alpha<2.$ One starts with a
measurable function $p:R\rightarrow(0,1)$ and a probability measure $\upsilon$
such that for $|x|>1,$%
\[
\upsilon(x)=\frac{[1-p(x)]dx}{2\gamma_{\alpha}|x|^{\alpha}}\text{ where
}\gamma_{\alpha}=\int\limits_{0}^{1}y^{\alpha-2}(1-\mathrm{e}^{-y})dy.
\]
We define now a stationary and reversible Markov chain, $(X_{n})_{n\in Z},$
with transition operator:
\[
Q(x,A)=p(x)\delta_{x}(A)+(1-p(x))\upsilon(A),
\]
where $\delta_{x}$ denotes the Dirac measure. It is stationary and ergodic
with the invariant distribution
\[
\pi(dx)=(\alpha-1)/(2|x|^{\alpha})dx\text{ for }|x|>1.
\]
Zhao et al. (2010) showed that $S_{n}=\sum\nolimits_{i=1}^{n}\mathrm{sign}%
(X_{i})$ does not satisfy the central limit theorem under any normalization.
In addition they showed that $\mathrm{var}(S_{n})\sim cn^{2/\alpha}.$ For
statistical inference of this example, we can use the CLT given in our Theorem
\ref{Theo1} immediately as $h_{n}n^{2/\alpha-1}\rightarrow0$.

\section{Conclusion and remarks}

In this paper we propose a method for constructing confidence intervals for
the mean or for testing statistical hypotheses for the mean of a dependent
stationary sequence with finite second moment. The method is robust in the
sense that we do not impose a specific restriction on the dependence structure
of the sequence except for the ergodicity and the consistency of the sample
mean in $L_{2}.$ The estimator we propose is $\hat{r}_{n},$ defined by
(\ref{def prime}) leading to the confidence intervals defined by
(\ref{conf int}). For applications, it is convenient to use a kernel $K$
following a standard normal distribution and to generate $(X_{i};1\leq i\leq
n)$ also from a standard normal variable. For this choice of $f$ and $K$, we
obtained $f(0)=1/\sqrt{2\pi}$, $\int K^{2}(u)du=1/(2\sqrt{\pi})$, $\int
u^{2}K(u)du=1$, $f^{\prime\prime}(0)=-1/\sqrt{2\pi}$. Thus, the plug in
estimator of the optimal bandwidth is%

\begin{equation}
h_{o}=(\frac{\overline{Y^{2}}}{n\sqrt{2}\bar{Y}^{2}})^{0.2},\text{ provided
}\mathrm{var}(\bar{Y}_{n})=o(n^{-0.8}). \label{sel h}%
\end{equation}
The $(1-\alpha)100\%$ confidence interval for $\mu_{Y}$ becomes
\begin{equation}
\left(  \hat{r}_{n}-z_{\alpha/2}\sqrt{\frac{1}{2h_{n}n^{2}}\sum_{i=1}^{n}%
Y_{i}^{2}},\text{ }\hat{r}_{n}+z_{\alpha/2}\sqrt{\frac{1}{2h_{n}n^{2}}%
\sum_{i=1}^{n}Y_{i}^{2}}\right)  , \label{conf int prac}%
\end{equation}
where%
\[
\hat{r}_{n}=\frac{1}{nh_{n}}\sum_{i=1}^{n}Y_{i}\exp[-(\frac{1}{h_{n}}%
X_{i})^{2}].
\]
It is easy to see that the size of the confidence interval depends on the
$\mathrm{var}(\bar{Y}_{n})$ via condition (\ref{rateP}), which restricts the
size of $nh_{n}.$ The larger $\mathrm{var}(\bar{Y}_{n}),$ the larger the size
of the interval.

Our result is asymptotic. We have conducted a numerical study to test the
performance of the confidence intervals based on formula (\ref{conf int prac})
on finite sample sizes. We have constructed confidence intervals based on
samples from an ARFIMA $(0,d,0)$ with innovations $(\xi_{j})_{j\in Z}.$ In our
simulations we vary the size of $d$, which controls the dependence strength,
and accordingly the size of $h_{n}$. Since the second moment of $Y$ is
important we also vary the distribution of $Y$ by considering various
distributions for the innovations. In all the situations, for relatively large
sample size, our methods returned reliable results.

Based on standard normal innovations we simulated an ARFIMA$(0,.09,0)$
sequence $(Y_{n}^{\prime})$ and set $Y_{n}=3+Y_{n}^{\prime}$. For a sample
size $n=100$, and using optimal bandwidth, we found that a $95\%$ confidence
for $\mu_{Y}$ is $(2.9,3.49),$ while for $n=1000$ the $95\%$ confidence for
$\mu_{Y}$ is $(2.81,3.04)$.

From $100$ confidence intervals constructed this way, $96$ of them covered the
real mean, which turns to be a statistically significant result for $95\%$
confidence intervals. Similar results were obtained for simulations based on
ARFIMA$(0,.09,0)$ with uniform innovations $U(-0.5,0.5)$ and with centered
$\chi^{2}(2)$ innovations.

When $d>0.1,$ according to (\ref{sel h}), $h_{o}$ does not satisfies the
restriction and we selected instead $h_{n}=n^{-2d}.$ For this case we
simulated samples from ARFIMA$(0,.49,0)$ and standard normal innovations,
$n=500.$ We obtained that from $100$ such simulations, $92$ of $90\%$
confidence intervals covered the mean.

\bigskip

\textbf{Acknowledgement}. The first author was supported by the College of
Liberal Arts Summer grant and the second author was partially supported by the
NSF grant DMS-1512936 and a grant from the Taft Research Center.

\end{document}